\documentclass[12pt]{article}
\usepackage[cp1251]{inputenc}
\usepackage{amssymb}
\usepackage{graphicx}

\begin{document}

\title{Zero-energy Andreev surface bound states in the lattice model}

\author{A.~M.~Bobkov\\ {\it Institute of Solid State Physics,}\\{\it Chernogolovka, Moscow
reg. 142432, Russia,}\\{\it e-mail: bobkov@issp.ac.ru} }

\maketitle

\begin{abstract} The conditions for zero-energy Andreev surface bound
states to exist are found for the lattice model of d-wave
superconductor with arbitrary surface orientation.  Both nearest
neighbors and next nearest neighbors models are considered. It is
shown that the results are very sensitive to the surface
orientation. In particular, for half-filled $(hl0)$-surface
zero-energy Andreev surface states only appear under the condition
that $h$ and $l$ are odd simultaneously.\end{abstract}

\vspace{0.5cm}

Significant feature of high temperature superconductors (HTS) are
zero-energy Andreev  surface bound states. The zero-energy states
(ZES) form on surfaces of a $d$-wave superconductor with
orientations different from the $(100)$, due to the sign change of
the order parameter. In high-temperature superconductors, such
states manifest themselves as the zero-bias conductance peak in
tunneling spectroscopy in the ab-plane
\cite{geerk88,hu94,tan95,mats95,buch95,tan952,cov96,xu96,fog97,bbs97,
cov97,alf97,ek97,ap98,alf981,alf982,sin98,wei98,apr99,deutsch99,ting110,
cov00,pairor02,greene02,wumou03,greene03}, the anomalous
temperature behavior of the Josephson critical
current~\cite{tan96,bbr96,ilichev01,blamire03} and the upturn in
the temperature dependence of the magnetic penetration
depth~\cite{walter98, bkk00,carr01}. At the same time the problem
of ZES at the surface with arbitrary orientation is still not
clear.

The conventional description of Andreev surface bound states, as
well as the most of inhomogeneous superconducting problems, is
based usually on the continuous quasiclassical approximation. From
the this viewpoint the conditions for ZES to exist are quite
simple. ZES are formed due changing of order parameter sign along
the quasiclassical trajectory. There are no ZES for $(100)$ (i.e.
$0^o$) orientation, and there are ZES at all values $k_{||}$ for
$(110)$ ($45^o$) orientation.  For intermediate surface
orientations the sign change does not take place for all incoming
momentum directions, and the weight of the ZES decreases with
deviation from $45^o$-orientation.

From the other side, the tight-binding BCS model is widely used
for theoretical description of HTS. This model gives the same (as
the continuous quasiclassical model) result for ZES at $(100)$ and
$(110)$ orientations. However, for the intermediate surface
orientations the question is very complicated. To the best of my
knowledge only the simplest orientations $(100)$, $(110)$, $(210)$
\cite{ting110,pairor02,tanaka99,hirsh04,walker99,tan00} have been
studied. And even numerical calculation couldn't give the general
answer for all surface orientations because of the lattice
specific.

In this paper the general analytical criterion for zero-energy
Andreev surface bound states to exist at the surface of arbitrary
orientation is presented. I consider two-dimensional tight-binding
model on square lattice. The surface orientation is assumed to be
arbitrary and characterized by the indexes $(hl0)$. Both nearest
neighbors and next nearest neighbors models are considered. For
simplicity I take the superconducting order parameter to be
spatially constant. This approximation is reasonable for studying
of low-energy quasiparticle states. The impenetrable surface is
assumed to be smooth.

The Hamiltonian for a pure singlet superconductor in the
tight-binding model can be written as:
\begin{eqnarray}
{\cal{H}} =- \sum_{{\bf x},{\bf x'},\sigma} t({\bf x},{\bf
x'})c^\dagger_{\sigma}({\bf x}) c_{\sigma}({\bf x'})+\nonumber \\
+ \sum_{{\bf x},{\bf x'}} \{ \Delta({\bf x},{\bf x'})
c^\dagger_{\uparrow}({\bf x}) c^\dagger_{\downarrow}({\bf x'}) +
h.c. \}, \label{ham}
\end{eqnarray}
Here $t({\bf x},{\bf x})=\mu$ is the chemical potential; $t({\bf
x},{\bf x}\pm {\bf a})=t({\bf x},{\bf x}\pm {\bf b})=t>0$, $t({\bf
x},{\bf x}\pm {\bf a}\pm {\bf b})=t'\le 0$ are the hopping
elements; $d$-wave superconducting pairing is defined for nearest
neighbors $\Delta({\bf x},{\bf x}\pm {\bf a})=-\Delta({\bf x},{\bf
x}\pm {\bf b})=\Delta$. Here ${\bf x}$ - are the positions of
lattice sites; ${\bf a,~b}$ -  are the basic lattice vectors. Then
the Bogoliubov-de Gennes equations take the form:
\begin{equation}
\sum\limits_{\bf x'}\left[ \begin{array}{cc}-t({\bf x},{\bf x'})&
\Delta ({\bf x},{\bf x'})\\ \Delta^*({\bf x},{\bf x'})&t({\bf
x},{\bf x'})\end{array}\right] {u({\bf x'}) \choose v({\bf x'})}=
E {u({\bf x}) \choose v({\bf x})}, \label{bdg}
\end{equation}

We define new coordinates ($\hat{x}$, $\hat{y}$), rotated with
respect to the crystal axes ($\hat{a}$, $\hat{b}$), where
$\hat{x}$ is the direction normal to the surface and $\hat{y}$ is
the direction along the surface. Superconductor is situated at
$x>0$. Lattice constant is taken to be unity, $a=1$. The system is
periodic along the $y$-direction with period $\sqrt{h^2+l^2}\equiv
d^{-1}$ and the crystal momentum component $k_y$ of a
quasiparticle is conserved. Instead of the usual square Brillouin
zone (BZ) $k_a \in [-\pi, \pi]$, $k_b \in [-\pi, \pi]$  we now use
the surface-adapted Brillouin zone (SABZ)\cite{pairor02,walker99}
given by $k_x \in[-\pi/d, \pi/d]$ and $k_y \in [-\pi d, \pi d]$.
Here $d =1/\sqrt{h^2+l^2}$ is the distance between the nearest
chains (layers) aligned along the surfaces, i.e. all
$x$-coordinates have discrete values with period $d$. The momenta
in the two coordinate systems are simply related through rotation
of an angle $\theta = \tan^{-1} h/l$.

Let us solve Eq.(\ref{bdg}) for half-space $x>0$ and fixed $k_y$.
General solution is constructed from all the solutions of uniform
problem, which don't grow at $x\to +\infty$. Then the wave
function for fixed $k_y$ can be written as
\begin{equation}
{u(x,k_y ) \choose v(x,k_y)}=\sum\limits_{\alpha}C_{\alpha}
{u_{\alpha}(k_y ) \choose v_{\alpha}(k_y)}~ e^{ik_{x,\alpha} x}
\label{wave} \enspace ,
\end{equation}
here summation should be taken over all solutions $k_{x,\alpha}$
of equation
\begin{equation}
E^2=\xi^2(k_x,k_y)+\Delta^2(k_x,k_y), \label{disp}
\end{equation}
with ${\rm Im}k_{x,\alpha}>0$. The boundary conditions are
\begin{equation}
{u(-jd,k_y ) \choose v(-jd,k_y)}=0,~~j=0,1,...,N-1, \label{bound}
\enspace ,
\end{equation}
where $N={\rm max}(h,l)$ for nearest neighbors model ($t\ne0$,
$t'=0$) or $N=h+l$ for next nearest neighbors model ($t,t'\ne0$)
\cite{hirsh04}. The total number of solutions (\ref{disp}) with
${\rm Im}k_{x,\alpha}>0$ equals to $2N$. Some of them correspond
to the intersections of line $k_y=const$ with Fermi-surface and
have small imaginary part of $k_{x,\alpha}$, the others correspond
to the point with $({\rm Re} k_{x,\alpha},k_y)$ far from
Fermi-surface. Therefore we obtain from (\ref{bound}) the system
of $2N$ liner equations for constants $C_{\alpha}$ with $E$ as a
parameter. Then the equality of the determinant of the system to
zero is the condition for existence of bound states with energy
$E$:
\begin{equation}
Det\left
(\begin{array}{ccc}u_1&...&u_{2N}\\v_1&...&v_{2N}\\
u_1 e^{ik_{x,1} d}&...&u_{2N} e^{ik_{x,2N} d}\\
...&~&...\\v_1 e^{ik_{x,1} (N-1)d}&...&v_{2N} e^{ik_{x,2N}
(N-1)d}\end{array} \right)=0. \label{det1}
\end{equation}

We only consider now the possibility for dispersionless states
with $E=0$ to exist in some region of $k_y$. Then all the
solutions of (\ref{disp}) have the form $(u_{\alpha}(k_y ),
v_{\alpha}(k_y))^T=(1,-i~\rho_{\alpha})$ with $\rho_{\alpha}=\pm
1$. From each point of intersection $k_y=const$ with Fermi-surface
we obtain one solution with $\rho_{\alpha}={\rm sgn}
(v_{f,x}(k_{x,f},k_y)\Delta(k_{x,f},k_y))$ in quasiclassical
approximation. And from each point far from Fermi-surface we
obtain two solutions with close values of $k_x$ and with opposite
values of $\rho_{\alpha}$.

Let $n$ and $m$ be numbers of solutions corresponding to
$\rho_{\alpha}=\pm1$, respectively. Then we can obtain after some
straightforward algebra that in the case of $n\ne m$
Eq.(\ref{det1}) is always true. For $n=m$ Eq.(\ref{det1}) can be
reduced to
\begin{equation}
\prod\limits_{\rho=-1,i<j}(k_{x,i}-k_{x,j})\prod
\limits_{\rho=+1,i<j}(k_{x,i}-k_{x,j})=0.
\label{det2}
\end{equation}

The wave vectors $k_{x,\alpha}$, corresponding to the same sign of
$\rho_{\alpha}$, can only coincide for a few values of $k_y$, for
which different parts or Fermi-surface intersect with each other.
Therefore we obtain the simple criterion for dispersionless
zero-energy bound states to exist: $n\ne m$. Since the solutions,
corresponding to the values of $\bf k$, which are situated far
from Fermi-surface, always appear in pairs with opposite signs of
$\rho_\alpha$, we can safely take into account only solutions with
$\bf k$ defined by the intersections of the line $k_y=const$ with
Fermi-surface.

Let's apply this criterion to the model under consideration. In
the quasiclassical approximation we need to obtain all
intersections of the line $k_y=const$ with Fermi-surface in SABZ,
and, then, calculate $\rho={\rm sign}~(v_{f,x}({\bf k})\Delta({\bf
k}))$ for all these points. Let us consider all values of $k_y$
simultaneously and find the positions of the edge of the regions
where zero-energy surface states exist.

Due to the symmetry of the normal metal quasiparticle energy
spectrum and the superconducting gap to the inversion: $\xi({\bf
k})=\xi(-{\bf k})$, $\Delta({\bf k})=\Delta(-{\bf k})$, we need to
consider only that points at Fermi-surface, where the sign of
$\rho$ changes. They are the points of the gap sign changing and
the points of $v_x$ sign changing.

It is easy to show that points of $v_x$ sign changing can't modify
the parameters $n$ and $m$. These points are the tangent points of
Fermi-surface and line $k_y=const$. On the one side (along
$k_y$-axis) from the point of $v_x$ sign changing there are two
solutions with opposite signs of $\rho_\alpha$. On the other side
from this point there are no real solutions, but there are two
solutions, which have large imaginary part of $k_x$ and opposite
signs of $\rho_\alpha$, too.

Thus let us consider only points of gap sign changing. We should
take into account univocal correspondence the SABZ and the usual
first BZ. Then there are only four points of gap sign changing in
SABZ, just as in BZ: $(k_a^0,k_b^0)=(\pm \nu \pi,\pm \nu \pi)$ in
crystal axes. Here parameter $\nu$ takes value $\pi^{-1}~{\rm
arccos}(-\mu/4t)$ for the nearest neighbors model and
\begin{equation}
\nu=\pi^{-1}~{\rm
arccos}\left(\frac{-\mu}{2(t+\sqrt{t^2-t'\mu})}\right) \label{p0}
\end{equation}
for more general case of the next nearest neighbors model.
Parameter $\nu$ is a relative coordinate of BZ point, where the
gap sign changing takes place and correlated with the filling of
the band. Maximal and minimal values of $\nu$ are 0 and 1. For the
simplest case of half-filling ($\mu=0$) we get $\nu=1/2$. But
$\nu$ is not strictly  the filling of the band.

Now we need to obtain the coordinates $k_y^0$ of the gap sign
changing points in SABZ. For $(hl0)$-orientation:
\begin{equation}
k_y^0=-k_a^0 \frac{l}{\sqrt{h^2+l^2}}+k_b^0
\frac{h}{\sqrt{h^2+l^2}},\label{ky}
\end{equation}
then for $k_y^0$-coordinates of four gap sign changing points
\begin{equation}
k_y^0=(\pm h \pm l)\pi\nu d.\label{ky1}
\end{equation}

Since $k_y$ is a crystal momentum, one can move the
$k_y^0$-coordinates of these points into the SABZ. Finally, we
obtain the following regions of $k_y$ where ZES exist:
\begin{equation}
|k_y^0|\in\left( k_{min},k_{max}\right),\label{kyinterval}
\end{equation}
where
\begin{equation}
k_{min}={\rm min}\{|F[(h-l)\pi\nu d]|,|F[(h+l)\pi\nu
d]|\}.\label{kmin}
\end{equation}
\begin{equation}
k_{max}={\rm max}\{|F[(h-l)\pi\nu d]|,|F[(h+l)\pi\nu
d]|\}.\label{kmax}
\end{equation}
Here $F[...]$ is a function, which shifts argument to the
permissible for SABZ value: $F[k_y]=\left(\{(1/2)+(k_y/2\pi
d)\}-(1/2)\right)2\pi d$, $\{...\}$-is a fractional part of the
argument. From Eq.(\ref{kyinterval}) we can see that the region of
$k_y$,where zero-energy bound states take place, always exists,
except for the case of
\begin{equation}
k_{min}=k_{max}.\label{except}
\end{equation}

\begin{figure}[tbh]
\begin{center}
\leavevmode
\includegraphics[width=1\columnwidth]{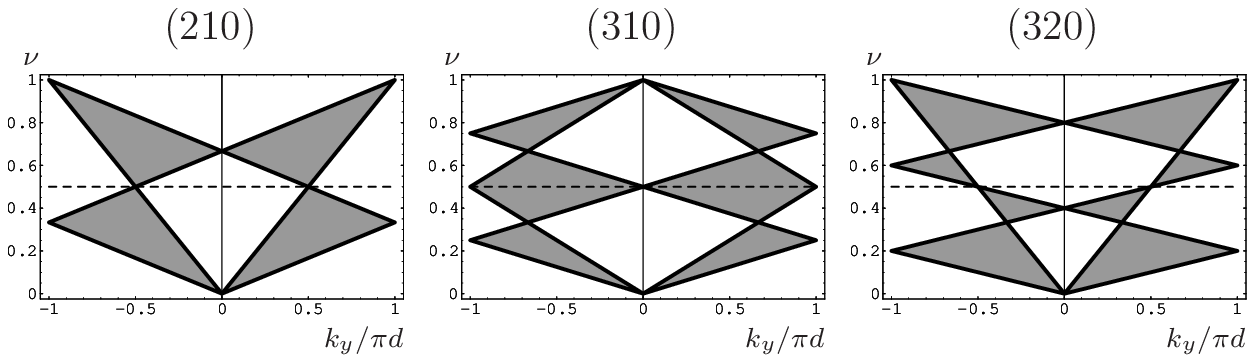}
\caption{Grey - regions of ZES existence in $(k_y,\nu)$ plane.
Black lines are the lines of zero gap. Dashed line corresponds to
half-filling $\mu=0$. Parameter $\nu=\pi^{-1}  {\rm
arccos}(-\mu/4t)$ for nearest neighbors model and $\nu=\pi^{-1}
{\rm arccos}(-\mu/2(t+\sqrt{t^2-t'\mu}))$ for next nearest
neighbors model.} \label{diagr}
\end{center}
\end{figure}

It is easy to obtain from Eqs.(\ref{kyinterval}-\ref{kmax}) the
regions of ZES existence for any cases under consideration. In
Fig.\ref{diagr} results for $(210)$, $(310)$, and $(320)$ surfaces
are shown. It is important to note, that regions with ZES and
regions without them are separated by the lines of zero gap (for
this values of $k_y$ superconducting gap vanishes for one of the
quasiparticle trajectories, forming the state).

For half-filled $(hl0)$-surface the result can be formulated in
the general form: zero-energy Andreev surface states appear only
for the case of odd $h$ and $l$. Moreover, it is seen from
(\ref{except}) that for any surface orientation one can find set
of values of $\nu$, for which there are no zero energy states.
From (\ref{except}) we have $h+l$ values of $\nu$ (and the same
number of band fillings):
\begin{equation}
\nu=i/l,~j/h,~~~i=1,...,h-1,~~j=1,...,l-1.\label{absence}
\end{equation}

\vskip 0.2cm {\it Conclusions.} The conditions for  zero-energy
Andreev surface bound state to exist are studied for the lattice
model of d-wave superconductor. Arbitrary surface orientation is
considered for nearest neighbors as well as for next nearest
neighbors models. The result is very sensitive to the surface
orientation, and doesn't change continuously under
surface-to-crystal angle rotation. In particular, for half-filled
$(hl0)$-surface zero-energy Andreev surface states only appear
under the condition that $h$ and $l$ are odd simultaneously.

\vskip 0.2cm {\it Acknowledgments.} I thanks I.~Bobkova and
Yu.S.~Barash for useful discussions. This work was supported by
the Russian Foundation for Basic Research under grant 02-02-16643
and Dynasty Foundation.

\end{document}